# A Physiologically-Based Flow Network Model for Hepatic Drug Elimination I: Regular Lattice Lobule Model

Vahid Rezania, Department of Physical Sciences, Grant MacEwan University, Edmonton, AB, Canada T5J 4S2
Rebeccah E. Marsh, Dept. of Physics, University of Alberta, Edmonton, AB, Canada T6G 2J1
Dennis Coombe, Computer Modelling Group Ltd., Calgary, AB T2L 2A6
Jack A. Tuszynski, Dept. of Physics and Experimental Oncology, University of Alberta, Edmonton, AB, Canada T6G 2J1

(Dated: October 25, 2011)

We develop a physiologically-based lattice model for the transport and metabolism of drugs in the functional unit of the liver, called the lobule. In contrast to earlier studies, we have emphasized the dominant role of convection in well-vascularized tissue with a given structure. Estimates of convective, diffusive and reaction contributions are given. We have compared drug concentration levels observed exiting the lobule with their predicted detailed distribution inside the lobule, assuming that most often the former is accessible information while the latter is not.

## I. INTRODUCTION

Pharmacokinetics aims to understand and predict behavior of various drugs through the body by studying the drug's distribution, absorption, metabolism and elimination from the body. Traditionally, such information can be extracted by examining drug concentration in the plasma or blood at several time points and generating a concentration-time curve that first rises based on the drug's absorption rate and then declines after reaching its maximum value governed by the drug's elimination rate. The result will be used to determine optimal dosing regimes, potential toxicities as well as possible drug-drug interactions.

To understand and predict drug behaviour through the body, compartmental models are commonly used in pharmacokinetics, Jacquez [1]. In general, each organ is represented by a compartment that contains a *homogenous* number of drug molecules undergoing a set of chemical kinetic reactions. The compartment's input/output is governed by *linear* kinetic processes with *constant* rate coefficients. At the end, the whole system that may contain several compartments is described by coupled first order differential equations whose solutions take the form of a sum of terms that are exponential in time.

Nonlinear drug-organ interactions, however, cannot be described adequately by classical compartment models. As observed in clinical data, the concentration-time curve shows a non-exponential time-dependency, asymptotically. Furthermore, observation of anomalous kinetics in the experimental data also suggests that the kinetic reactions should be occurring on or within fractal media with a time-dependent kinetic rate coefficients (Anacker and Kopelman [2], Kopelman [3]).

To address the above shortcomings several explanations have been proposed, including a stochastic random walk model for the drug molecules (Wise et al [4]), convective-diffusive transit behaviour in the liver (Norwich and Siu [5]), gamma-distributed drug residence time (Wise [6]), transient fractal kinetics (Anacker and Kopelman [2]), and fractal Michaelis-Menten kinetics (Marsh and Tuszynski [7]). All of these studies, however, represented an organ as a homogenous compartment that is not realistic, i.e. physiologically-based.

### A. Lattice Models of the Liver

Alternatively, to take into account the organ heterogeneity and simulate enzyme kinetics in disordered media, lattice models have been introduced by investigators. Berry [8] performed Monte Carlo simulations of a Michaelis-Menten reaction on a two-dimensional lattice with a varying density of obstacles to silate the barriers to diffusion caused by biological membranes. He found that fractal kinetics resulted at high obstacle concentrations. Kosmidis et. al [9] performed Monte Carlo simulations of a Michaelis-Menten enzymatic reaction on a two-dimensional percolation lattice at criticality. They found that fractal kinetics emerged at large times.

Previously [10], we developed a network model of the liver consisting of a square lattice of vascular bonds connecting two types of sites that represent either sinusoids or hepatocytes. Random walkers explored the lattice at a constant velocity and were removed with a set probability from hepatocyte sites. To simulate different pathological states of the liver, random sinusoid or hepatocyte sites were removed. For a lattice with regular geometry, it was found that the number of walkers decayed according to an exponential relationship. For a percolation lattice with a fraction p of the bonds removed, the decay was found to be exponential for high trap concentrations but transitioned to a stretched exponential at low trap concentrations.

The models described above are all basic random walk models, and the lattices are abstract representations of the geometry of the space. The objective of this paper is to develop a lattice model that incorporates realistic anatomical and physiological properties of the liver as well as the flow of blood with reacting tracers

In this paper, we introduce a physiologically-based lattice model of the functional unit of the liver that takes into account parameters such as the distribution volume, permeability, and porosity of the liver vasculature and cells. Instead of simple random walkers, a blood-like fluid containing drug molecules flows through the vascular space of the lattice. This model takes into account flow-limited and diffusion-limited exchange of drug molecules into the extravascular space from the sinusoids; sequestration of the drug molecules within liver cells with enzymatic transformation; and exchange of the metabolized drug molecules back from liver cells to the vasculature. Estimates and consequences of the competing flow processes are given. Furthermore, the enzymatic transformation of the drug can be either simple or saturable. The model allows us to include the effects of the intrahepatic mixing process on the enzymatic transformation of drug molecules.

## II. THE LIVER AND DRUG KINETICS

### A. Liver Architecture

At the macroscopic level, the liver consists of three vascular trees, two supply trees that originate from the portal artery and hepatic vein, and one collecting tree that drains into the portal vein [11]. The vessels bifurcate down to the terminal arterioles and venules, which are organized into portal tracts along with a terminal bile duct. Liver cells, called hepatocytes, radiate outward from the terminal vessels. These plates of hepatocytes are interspersed by sinusoids, which play the role of the capillary in the liver, and the spaces of Disse, which are the extravascular space of the liver [12]. Finally, the blood is collected and removed by the hepatic venules.

### B. Functional Unit

The functional unit of an organ is the smallest structural unit that can independently serve all of the organ's functions [13]. Because of its complexity, there is continued debate about what the functional unit of the liver should be. The classic lobule is a hexagonal cylinder, centered around a hepatic venule and with portal tracts situated at the corners. The portal lobule has a similar shape but is centered about a portal tract with the hepatic venules at the periphery [14]. The acinus is another proposed unit and is based on the pattern formed by the cords of hepatocytes between two central venules. Matsumoto and Kawakami [15] suggested that the classic lobule can be divided into primary lobules, which are cone-shaped and each fed by one portal tract and drained by one hepatic venule. Teutsch and colleagues [16, 17] performed a morphological study of rat and human liver lobules, and their results support the idea of a secondary unit made up of primary units in what they term as a modular architecture.

They conclude, however, that the primary unit is more polyhedral in shape than conical. Other experiments done by Ruijter et al. [18] suggest that the primary unit is needle-shaped and that there are equal amounts of portal and central vein associated with one unit. For this study, the primary unit is taken to be one-fourth of the classical lobule. The relevant anatomical values are listed in Table I.

**C. Elimination Kinetics**

Here drug uptake and elimination (i.e. conversion to metabolized product) is viewed as a single-step saturable process following Michaelis-Menten kinetics [19], such that

$$\frac{dC(t)}{dt} = v_{max} C(t)/(K_m + C(t)) \qquad (1)$$

Here C(t) is the local drug concentration and $\frac{dC(t)}{dt}$ is the drug metabolization rate. (In what follows, C(t) is expressed as $\rho x_i$ with $\rho$ the fluid density and $x_i$ mole fractions of i-th species). Note we are explicitly modeling the drug transport to an individual hepatocyte surface via our lobule lattice model and assume an effective one-step reaction transformation beyond that point. We recognize that drug incorporation and elimination is still a multi-step process even once the drug reaches the cell surface however. It is hoped that these approximations ignore processes occurring on a shorter time scale than the experimental resolution. Nonetheless, we explore possible complications via simple sensitivities to the choice of reaction time constant.

## III. MODEL AND METHOD

The primary unit of the liver was approximated by a symmetry element of a 51 x 51 square lattice such that four units make up one lobule. The architecture of the lattice consists of hepatocyte grid cells interlaced by a network of narrower sinusoidal grid cells (Figure 1). The diameter of the sinusoid grid cells was taken to be 0.0006 cm, and the diameter of the hepatocyte grid cells was taken to be 0.0024 cm. The length of the lattice is thus 0.0750 cm per side. Doubling this value gives a lobule diameter of 0.150 cm, which is consistent with values listed in Table I.

Convective molar flux is modelled according to Darcy's Law [20]:

$$J^c_{ik} = \rho\ x_i\ v_k = \rho\ x_i \frac{K_k}{\mu} \nabla_k p \qquad (2)$$

where $J^c_{ik}$ is the i-th component of fluid flux in k-direction, $\rho$ and $\mu$ are fluid molar density and viscosity, $v_k$, $K_k$ and $\nabla_k p$ are the Darcy velocity, permeability, and pressure gradient in direction k, respectively. The blood viscosity $\mu$ is taken to be 3.5 mPa-s (3.5 centipoise). Blood molar density is assumed that of water, $\rho$ at 55.4 moles/cm$^3$. It is emphasized that in this paper, following Darcy's Law and the conventions of flow in porous media, permeability K is defined as a measure of the transmissibility of a grid cell to the flow of a fluid, and is expressed in units of area (e.g. cm$^2$).

Each sinusoid grid cell represents a tubular vessel of diameter 2a. Taking the ratio of the volume of the vessel to the volume of the grid cell yields a porosity of

$$\phi_{sin} = \frac{\pi a^2 a}{a^3} = \frac{\pi}{4} = 0.7854 \qquad (3)$$

For a cylindrical tube, the permeability can be calculated from the tube radius [13]

$$K_{sin} = \frac{a^2}{8} = 1.126\ \text{micron}^2 \qquad (4)$$

Here and in what follows, it is noted that porosity is dimensionless while permeability has units of a characteristic length squared.

Each parenchymal grid cell represents a cellular (hepatocyte) component and an extracellular (space of Disse) component. A ratio of 0.75 to 0.25 was chosen for their respective contributions to the volume, and the porosity of the parenchymal sites was therefore 0.25.

The corresponding permeability of the tissue grid cells is estimated from a Carmen-Kozeny model [20] of flow around a spherical object (the hepatocyte) of radius 12 microns. This states the permeability is proportional to the object's diameter $D_p$ and the porosity $\phi_{tis}$ as follows:

$$K_{tis} = \frac{D_p^2 \phi_{tis}^3}{180(1-\phi_{tis})^2} \tag{5}$$

An ideal result where $D_p/L=1$ or $R/L=0.5$ with L as grid size and R as particle radius gives

$$(1-\phi_{tis}) = \frac{4\pi(0.5)^3}{3} = 0.5236 \quad \text{or} \quad \phi_{tis} = 0.4764 \tag{6}$$

Then assuming as we do, $L = D_p = 24$ microns, the Carmen-Kozeny formula yields

$$K_{tis} = 1.262 \text{ micron}^2$$

The Carman-Kozeny expression basically states physically that the order of magnitude of the permeability scales with the particle size squared.

The above analysis is an ideal result as the ECM around the cell particle will further reduce porosity. Thus we could study a range of tissue effective porosities, leading to a range of effective tissue permeabilities. Table II summarizes a range of possible values. In this paper, we will utilize the "base case" value for tissue porosity and permeability, while the effects of more extreme choices will be examined in a second paper.

The convective driving force originates from an input site corresponding to a terminal portal venule at one corner of the lattice and an output site corresponding to a terminal hepatic venule at the opposite corner. For simplicity, the hepatic artery blood supply, which is lower in volume and pulsatile in nature, is omitted for the current simulations. The pressure value at the inlet and the outlet are taken to be $P_{in} = 103$ kPa and $P_{out} = 101.8$ kPa, respectively. After subtracting the atmospheric pressure, these values are consistent with experimental values quoted by Rappaport [21], who found that the terminal portal venule pressure was in the range 0.59 kPa to 2.45 kPa and that the terminal hepatic venule pressure was 0.49 kPa. As we shall demonstrate, this applied pressure differential results in a convective flow level that is determined primarily by the effective permeability of the lobule. Thus various liver damage scenarios can be expected to affect this flow. This aspect of the modelling is of practical importance and will be explored in more detail in a separate publication.

For multicomponent flow, the model tracks the compositions (molar or mass fractions) of all components in the fluid. In addition to convective transport a diffusive flux contribution of

$$J_{ik}^d = D_{ik} \nabla_k (\rho x_i) \tag{7}$$

is considered with $J_{ik}^d$ being the molar diffusive flux and $D_{ik}$ the diffusion constant of species i in direction k. The estimated diffusion constant in all directions used here is based on a molecular weight rescaling of glucose diffusion. Here

$$D(GLC) = 7.1 \times 10^{-10} \frac{m^2}{\sec} \tag{8}$$

With a cubic root of the molecular weight ratio of glucose to PAC used as conversion factor (Factor = $(180/854)^{0.33}$ = $(1/4.74)^{0.33}$ = 1/1.68), an estimated effective diffusion constant for PAC is

$$D(PAC) = \frac{7.1 \times 10^{-10}}{1.68} = 4.2 \times 10^{-10} \frac{m^2}{\sec} \quad (9)$$

Tissue effective diffusion value should be less, here we employ an order of magnitude reduction in the value of D

$$D(PAC) = \frac{7.1 \times 10^{-11}}{1.68} = 4.2 \times 10^{-11} \frac{m^2}{\sec} \quad (10)$$

Effective diffusion constants for PAC-OH are assumed identical to PAC values. These values are converted to the simulation units of $cm^2$/min and also summarized in Table II.

The drug paclitaxel was used as a reactive tracer, and its Phase I metabolism was modeled using the general formula of one paclitaxel (PAC) molecule being transformed into the metabolite 6ff -hydroxypaclitaxel (PAC-OH) by the cytochrome P450 (CYP) isozyme CYP2C8 [22].

$$PAC + CYP \; > \; PAC\text{-}OH + CYP \quad (11)$$

The enzyme only exists in grid cells containing hepatocytes so all reaction is localized in these sites. In this paper, saturable Michaelis-Menten kinetics are assumed, defined by a maximum rate $v_{max}$ (in units of molar fraction/min) and a half saturation value $K_M$ (in units of molar fraction). This reaction proceeds in a linear manner at a rate characterized by $k = v_{max}/K_m$ (in units of $min^{-1}$) when injection concentrations are much below the half saturation value.

Reaction parameter values are based on the work of Vaclavikova et al [23] who measured directly PAC conversion to PAC-OH kinetics without any tissue distribution issues and uptake by the cell itself, as they use microsomes directly as the source of CYP. As such, any bottlenecks associated with drug uptake should imply that reaction rates would be slower than those based on parameters values given by Vaclakova et al [23]. We are modeling tissue distribution effects separately based on our lobule model. Table III summarizes the reaction parameters.

The simulations were performed using the STARS advanced process simulator designed by the Computer Modelling Group (CMG) Ltd. in Calgary, Alberta, to model the flow and reactions of multiphase, multicomponent fluids through porous media [24], [25], [26]. Additionally, STARS has earlier been used to model reactive flow processes in cortical bone [27], [28], [29], [30] as well as through the intervertabral disk [31].

## IV. RESULTS

**Non-reactive Flow Characteristics**

As discussed above, flow is induced on the regular lattice of Figure 1 by applying a pressure difference across the inlet and outlet points. With the chosen lobule flow parameters for porosity, permeability, and blood viscosity, this translates to a steady flow rate 2.1 $cm^3$/min as illustrated in Figure 2. A short timescale of about $2.0 \times 10^{-5}$ min needed to establish this pressure gradient is also illustrated in this figure by expanding the time axis. Figure 3 demonstrates the steady state velocity profile throughout the lattice, illustrating both the diverging/converging nature of the flow near the inlet and outlet points, as well as the orders of magnitude difference of the flows in the sinusoids and tissues, respectively. (This plot uses a logarithmic colour scale axis).

When blood with a relative composition of 1 micro-gram paclitaxel ($1.8 \times 10^{-8}$ mole fraction) is infused into the lattice assuming nonreactive hepatocytes, the time required to traverse the lattice is approximately 1 min as demonstrated in Figure 4. This production profile is convective flow dominated as the addition of diffusion minimally alters the production profile.

The evolution of the paclitaxel concentration on the lattice followed a spatially homogeneous progression (Figure 5), which shows the increasing levels of injected drug after 0.01 min and 0.14 min. By 0.14 min, paclitaxel is being seen at the outlet of the lobule. After 0.5 min, paclitaxel completely covers the lattice (not shown).

If the diffusive flow contribution is removed, however, the paclitaxel profiles on the lattice are significantly different. As is also illustrated in Figure 5, again at 0.01 min and 0.14 min, a distinct two-scale behavior is noted, whereby the sinusoids are first infused with the drug, and only at later times do the drug levels in the tissue approach injected concentration levels. This behavior reflects the convective levels of flow in the sinusoids and tissues noted earlier (Figure 3). A further comparison of Figure 5 cases reveals that the sinusoid drug concentration levels in the two cases are similar, however, explaining the similar drug production characteristics note in Figure 4, as paclitaxel is produced directly from the sinusoids.

**Base Case Reactive Flows**

The effects of paclitaxel drug metabolism by hepatocytes are next considered. Here the base case reaction parameters of Table III are employed, and the same injected paclitaxel injected concentration (1.8e-8 mole fraction) is considered. With the employed reaction half saturation constant value of $1.8 \times 10^{-7}$ mole fraction, this injection level implies the Michaelis-Menten model reduces to a linear reaction scheme.

Figure 6 illustrates injected drug and produced drug and metabolite production for this case. Essentially at this reaction rate, all injected paclitaxel is converted to metabolite by the lobule hepatocytes. The production profile of PAC-OH here is identical to the production profile of PAC in the non-reaction case, as shown in Figure 4. Figure 7 shows the PAC and PAC-OH profiles across the lobule lattice at 0.01 min, 0.14 min, and 0.50 min, respectively. The PAC concentrations in the sinusoids and the PAC-OH concentrations in the tissue are equivalent to the PAC concentrations in both sinusoids and tissue for the non-reacted case (Figure 5). Figure 7 also shows most clearly there is an inlet distance over which the reaction conversion time is not fast enough to convert the injected paclitaxel.

Figure 8 illustrates injected drug and produced drug and metabolite production for the same case except that diffusive transport has been removed. In contrast to Figure 6 with diffusion, there is now only a limited amount of conversion of PAC to PAC-OH even at long times. The PAC and PAC-OH profiles at various times (0.01 min, 0.14 min, and 0.50 min) as shown in Figure 9, confirm this behavior where it is shown that the PAC concentration in the sinusoids propagates throughout the lobule, while the PAC-OH concentrations in the tissue increase less rapidly and up to a lower level.

**Reactive Flow Sensitivities**

In this section, the consequences of the chosen reaction parameters are illustrated. Figure 10 shows production behavior with a 100-fold reduction in maximum reaction rate and with diffusion effects included. The small level of produced PAC indicates that reaction rates must be reduced to about this level before any significant change in drug production behavior can be expected. Drug distribution in the lobule for this case is shown in Figure 11 for the times 0.01 min, 0.14 min, and 0.50 min. This figure should be contrasted with Figure 9. Here the early time results and upstream results for PAC distributions at longer times are quite different, reflecting the reduced reaction rate. However, the later time and downstream results for PAC-OH distribution resemble quite closely the faster reaction limit. Here the propagation time across the lobule gives enough time to compensate for changes in reaction rate. In summary, reaction rates larger than the base case or even 10-fold reduction from base case can be expected to produce very similar drug production behavior and differences only in the inlet region of the lobule are to be envisioned.

In contrast, once a critical time-scale is crossed, much more significant changes in drug distribution behavior can be expected, both internally throughout the lobule and in terms of produced profiles. Figure 12 shows drug metabolite production behavior with a 1000-fold reduced metabolic rate, and including diffusive mixing. Here almost equal levels of PAC and PAC-OH are seen exiting the lobule. At 0.01 min almost no PAC-OH is converted in the lobule at this rate (see Figure 13), while at later times (0.14 min and 0.50 min), converted PAC-OH starts to be seen at the outlet regions at levels similar to PAC. Essentially, the inlet region behavior occurring at faster reaction rates now covers the whole lobule region.

Finally, sensitivities to injected PAC concentrations were explored, utilizing injection concentrations of $1.8 \times 10^{-7}$ and $1.8 \times 10^{-6}$ mole fractions (i.e. clearly above that of the base case $1.8 \times 10^{-8}$). In these runs, the base case reaction parameters were maintained. In particular, the half saturation value of $1.8 \times 10^{-7}$ was employed, indicating that the linear, intermediate, and saturation levels of the Michaelis-Menten expression were being probed with the three injected concentration levels. As illustrated in Figure 14 for the runs without diffusion, the production profiles of PAC-OH remained unchanged for each case, as long as the production maxima were rescaled to the corresponding injection concentrations. Apparently, with fast reaction rates, the Michaelis-Menten form had little impact on production behavior.

## V. ANALYSIS

The results we have presented can be rationalized by a comparison of process timescales.

Calculation of breakthrough times can be based on two concepts: either only the sinusoids are accessible or the whole lobule (tissue+sinusoid) is accessible to injected species. The sinusoid pore volume in our element is $4.77 \times 10^{-7}$ cm$^3$ while the complete lobule element volume is $7.344 \times 10^{-7}$ cm$^3$. Because the steady state flow rate in our model is $2.44 \times 10^{-6}$ cm$^3$/min (see Figure 2), this means the breakthrough time is $4.77/24.4 = 0.20$ min for just sinusoid accessibility and $7.34/24.4 = 0.30$ min for the whole lobule-sampled space. These should be viewed as two limiting vertical lines on the time history plots as two ideal limits without any diffusion or mixing effects (physical or numerical). It is noted for example that our time history plot of PAC production with no reaction and with or without physical diffusion (see Figure 4) has a produced concentration of $0.9 \times 10^{-8}$ (i.e. half of the injected $1.8 \times 10^{-8}$ concentration) at 0.19 min, about what is expected. The main point here is that most of the production behavior differences for our various cases should lie between these two ideal "half-value" limits.

The next timescale is governed by a "pressure diffusion" coefficient

$$D_{pres} = \frac{K}{\phi \mu C_p^{eff}} \tag{12}$$

Here $C_p^{eff}$ is an effective compressibility accounting for both fluid and tissue structure effects. Fluid (water compressibility) is of the order of 5e-7 kPa$^{-1}$. For liver (soft tissue) structural compressibility, we have chosen $1.8 \times 10^{-5}$ kPa$^{-1}$. Using these choices and the base case parameters of this study from Table II, we obtain

$$D_{pres} = 1.5 \times 10^{+4} \frac{cm^2}{\min} \tag{13}$$

This parameter essentially describes the time taken for pressure to come to a steady state distribution as follows. Utilizing a characteristic distance $d = 0.15$ cm (the lobule element size for the pressure calculation), this time is then

$$T_{pres} = \frac{d^2}{D_{pres}} = 1.5 \times 10^{-5} \min \tag{14}$$

Figure 2 also illustrates this characteristic time. This timescale is essentially a function of fluid properties and lobule structure (through $\phi$ and K). If we were to consider pulsatile flow effects caused by hepatic artery inflow, this timescale would be much more important to the general process description and a more precise definition of compressibility might be warranted. For the present, these numbers just indicate that steady state pressure is achieved more quickly than other process effects.

The third timescale is determined by particle diffusion. Here we have chosen diffusion constants based on paclitaxel size and simple estimates of tortuosity. The parameter choices used here are $D_{sin} = 2.5 \times 10^{-4}$ cm$^2$/min and $D_{tis}$ = 2.5 x10$^{-5}$ cm$^2$/min. Again with a choice of characteristic distance d = 10 micron = 1 x10$^{-4}$ cm, the times required for particles to diffuse are

$$T_{diff(\sin)} = \frac{d^2}{D_{\sin}} = 4.0 \times 10^{-3} \text{ min} \qquad (15)$$

$$T_{diff(tis)} = \frac{d^2}{D_{tis}} = 4.0 \times 10^{-4} \text{ min} \qquad (16)$$

Our diffusion values should be viewed as highly optimistic. In particular, pactlitaxel is normally not molecularly dissolved, but rather it is some type of micellar complex with Cremophor EL surfactant, so the effective diffusion coefficient for this complex is probably one or more orders of magnitude smaller than what has been estimated. Thus the limits of diffusion and non-diffusion cases are meaningful extremes of what might be expected, for small molecules and large nanoparticles, respectively.

The final timescale is reaction rate. The base limiting reaction rate 6 x10$^{-3}$ min$^{-1}$ converts to a reaction time (reaction half-life) of

$$T_{reac} = \frac{\ln(0.5)}{6 \times 10^{-3}} = 1.0 \times 10^{+2} \text{ min} \qquad (17)$$

This is essentially seen as instantaneous compared to the other timescales considered. Reducing this basic reaction rate by a factor of 100 or 1000 causes the reaction process to be more similar to the other timescales and different production profiles of PAC and PAC-OH result, as has been shown.

Table IV summarizes the relevant assumed timescales.

## VI. CONCLUSION

In pharmacokinetics, lattice models are introduced to address the non-heterogeneity of the organs on the drug distribution that has a significant impact on drug propagation throughout the body as shown by analysis of clinical data.

Here we utilize the interpretation of the liver as an ensemble of islands of metabolic activity and focus on the liver lobule itself. In contrast to most earlier studies that assumed that drug molecules only randomly propagate through the system, we have emphasized the dominant role of convection in well-vascularized tissue with a given structure. We have utilized an idealized representation to analyze the factors affecting drug propagation and metabolism. The lobule is divided into hepatocyte cells that are interlaced with narrower sinusiodal grid cells. These cells are connected by constant permeability throughout the entire system. The drug molecules convectively flow through the sinusiodal along with blood (water here) and diffuse to the hepatocyte where metabolisms are taking place. A sensitivity analysis of convective, diffusive and reaction parameters is performed and estimates of their contributions are presented. We have compared the drug concentration levels observed exiting the lobule with their predicted detailed distribution inside the lobule, assuming that most often the former is accessible information while the latter is not.

This is the first paper of series of papers on physiologically-based lattice models for liver. In this paper, we consider an idealized lobule lattice in order to understand the basic functionality of the unit and underlying mechanisms through simulations and also to set a basis for future studies. A following paper will expand the analysis to include sensitivities associated with variations in lobule structure, which could reflect extents of liver damage.

**Acknowledgements**

J.A.T. acknowledges funding support for this project from NSERC. The Allard Foundation and the Alberta Advanced Education and Technology.

## TABLES:

**TABLE I: Anatomical Parameter Values for the Liver [11], [12]**

| Parameter | Value |
|---|---|
| Hepatocyte diameter | 12-24 $\mu$m |
| Diameter of liver cell sheets | 25 $\mu$m |
| Lobule diameter | 1 - 2.5 mm |
| Mean sinusoid diameter | 7.3 $\mu$m |
| Vascular tissue component | 28 - 30% |
| Specific gravity of liver | 1.05 |
| Liver volume | 1071 $\pm$ 228 cm$^3$ |

**TABLE II: Lobule Regular Lattice Flow Parameters**

| Parameter | Characteristic Unit | STARS Unit |
|---|---|---|
| Sinusoid Grid Cell Size | 6 micron | 0.0006 cm |
| Sinusoid Porosity | 0.7854 | 0.7854 |
| Sinusoid Permeability | 1.125 micron$^2$ | 1.140 Darcy |
| Sinusoid Effective Diffusion | 4.2e-10 m$^2$/sec | 2.5 x10$^{-4}$ cm$^2$/min |
| Tissue Grid Cell Size | 24 micron | 0.0024 cm |
| Tissue Porosity (Ideal) | 0.4764 | 0.4764 |
| Tissue Permeability (Ideal) | 1.230 micron$^2$ | 1.246 Darcy |
| Tissue Porosity (Base) | 0.2382 | 0.2382 |
| Tissue Permeability (Base) | 7.35 x10$^{-2}$ micron$^2$ | 7.45x10$^{-2}$ Darcy |
| Tissue Porosity (ECM) | 0.1191 | 0.1192 |
| Tissue Permeability (ECM) | 6.883 x10$^{-2}$ micron$^2$ | 6.97 x10$^{-2}$ Darcy |
| Tissue Effective Diffusion | 4.2 x10$^{-11}$ m$^2$/sec | 2.5 x10$^{-5}$ cm$^2$/min |

(1 Darcy = 0.9869 microns**2 in engineering permeability units).

**TABLE III: Paclitaxel Kinetic Elimination Michaelis-Menten Parameters (converted* from Vaclavikova et al, their Table 4)**

| Parameter | Characteristic Unit | STARS Unit |
|---|---|---|
| Maximum Rate Vmax | 0.06 micronM/min | 1.08 x10$^{-9}$ molefrac/min |
| Half Saturation Constant Km | 10.0 micronM | 1.8 x10$^{-7}$ molfrac |
| Linear Rate Vmax/Km | 6.0e min$^{-1}$ | 6.0 x10$^{-3}$ min$^{-1}$ |

*Their Table 4 quotes Vmax = 61 picomol/mg_protein/min. Using their microsomal protein concentration of 1 mg/ml, these numbers convert to Vmax = 0.06 micronM/min.

**TABLE IV: Assumed Lobule Process Time Constants**

| Process | Time |
|---|---|
| Convective Transit Time (sinusoid network only) | 0.200 min |
| Pressure Relaxation Time Constant (in sinusoids) | 1.5 x10$^{-5}$ min |
| Diffusion Relaxation Time Constant (CYP/CYP-OH in sinusoids) | 4.0 x10$^{-3}$ min |
| Diffusion Relaxation Time Constant (CYP/CYP-OH in tissue) | 4..0 x10$^{-4}$ min |
| Base Case Metabolic Uptake/Elimination Time Constant | 1.0 x10$^{+2}$ min |

# FIGURES

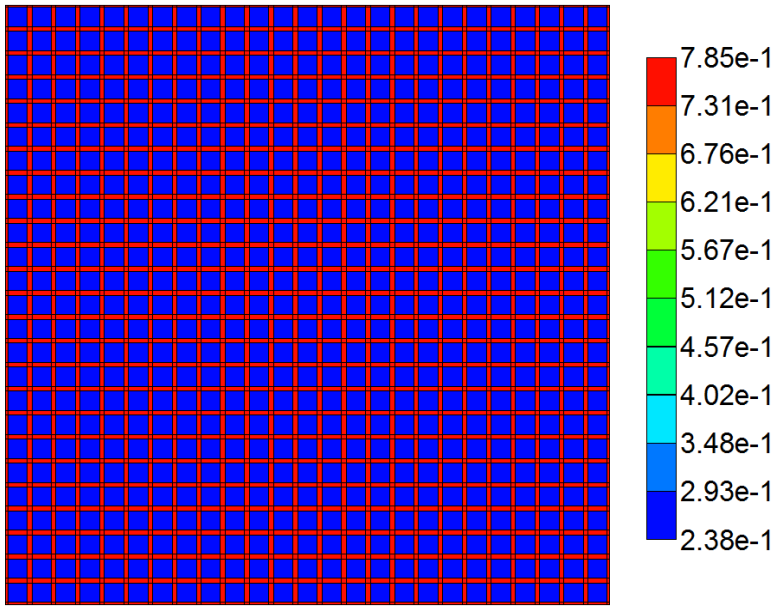

Figure 1: Homogeneous lattice. The high porosity bands represent sinusoids and the lower porosity regions represent tissue containing hepatocytes.

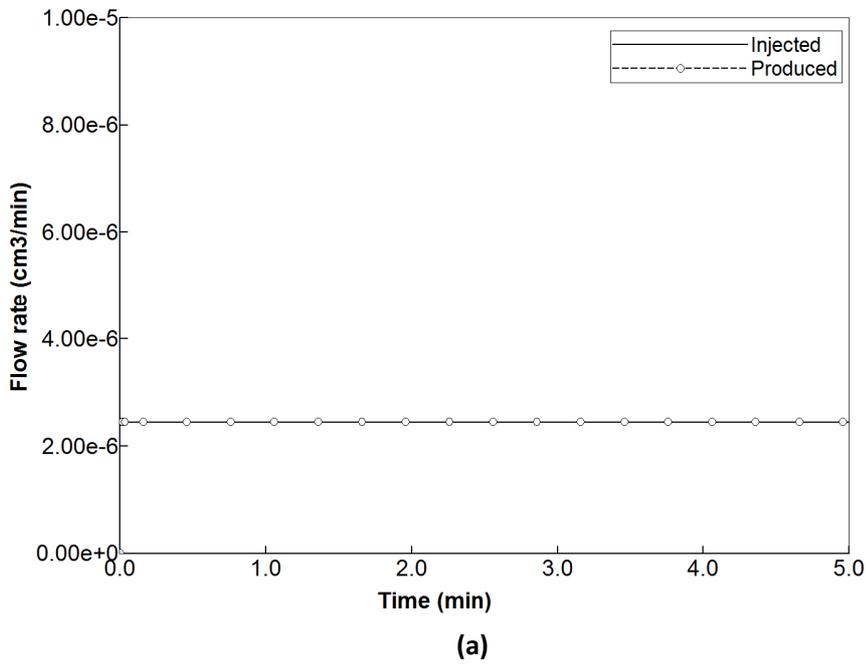

(a)

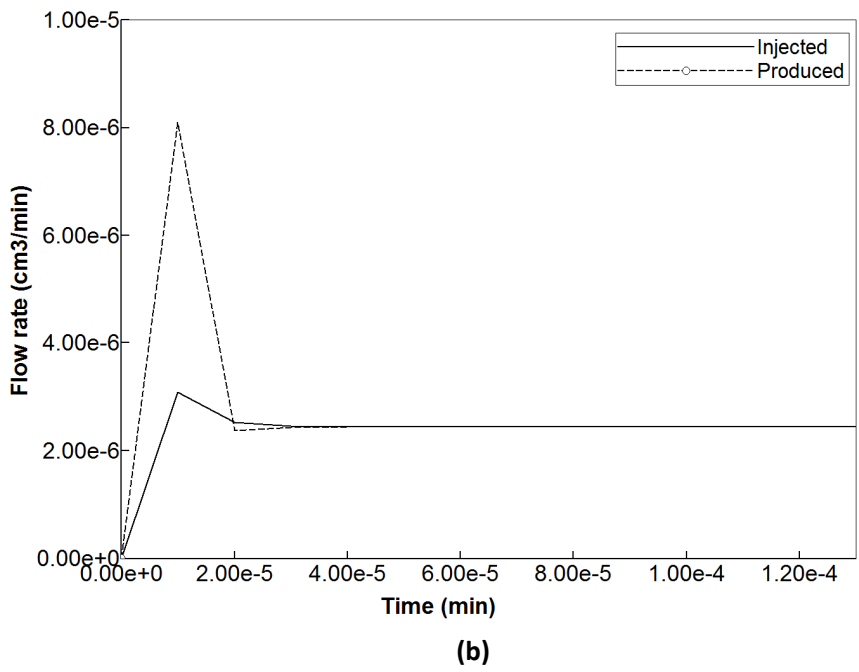

(b)

**Figure 2: (a) Steady state flow across the lobule, (b) Short time flow transients.**

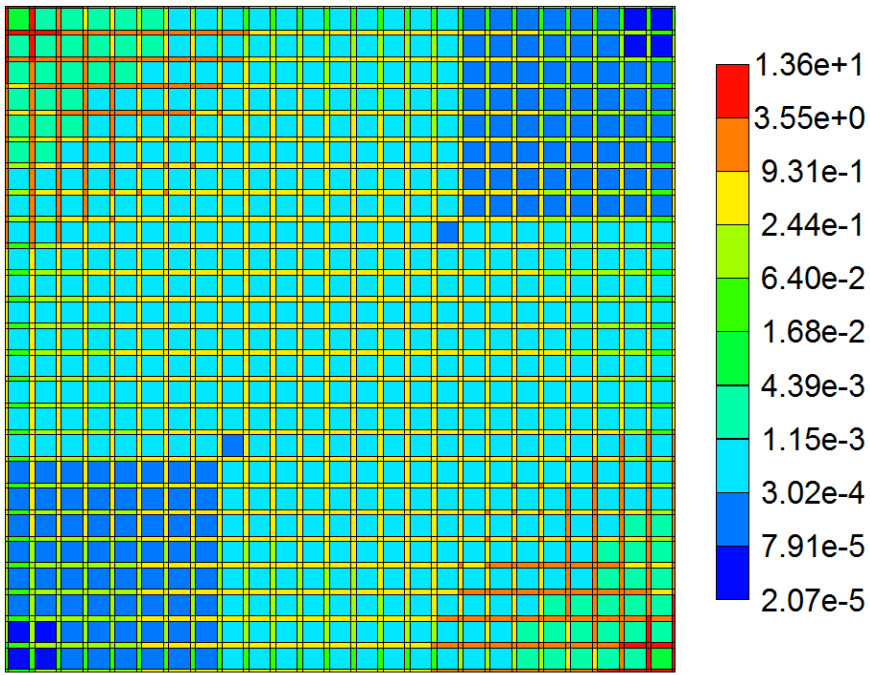

Figure 3: Steady state velocity profile across the lobule.

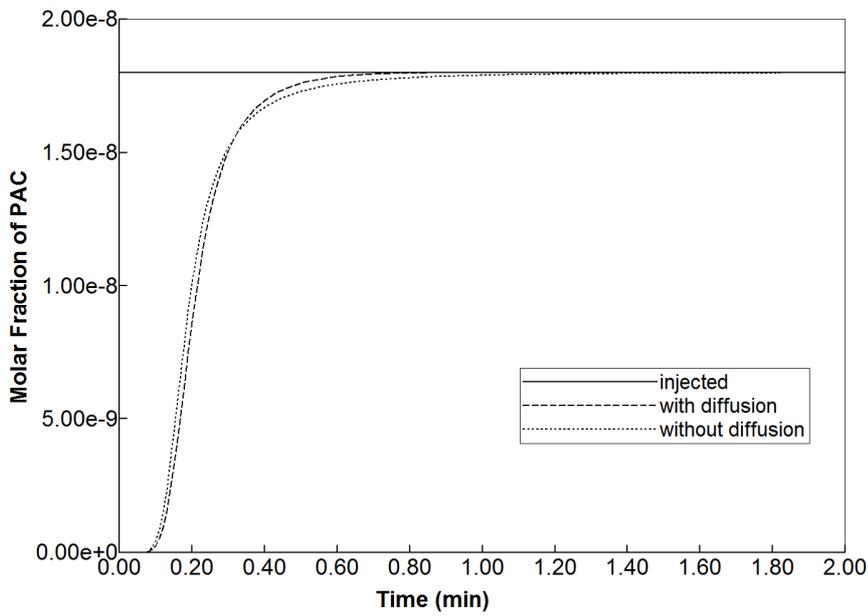

Figure 4: Non-reactive PAC drug propagation across the lobule, with and without diffusion effects.

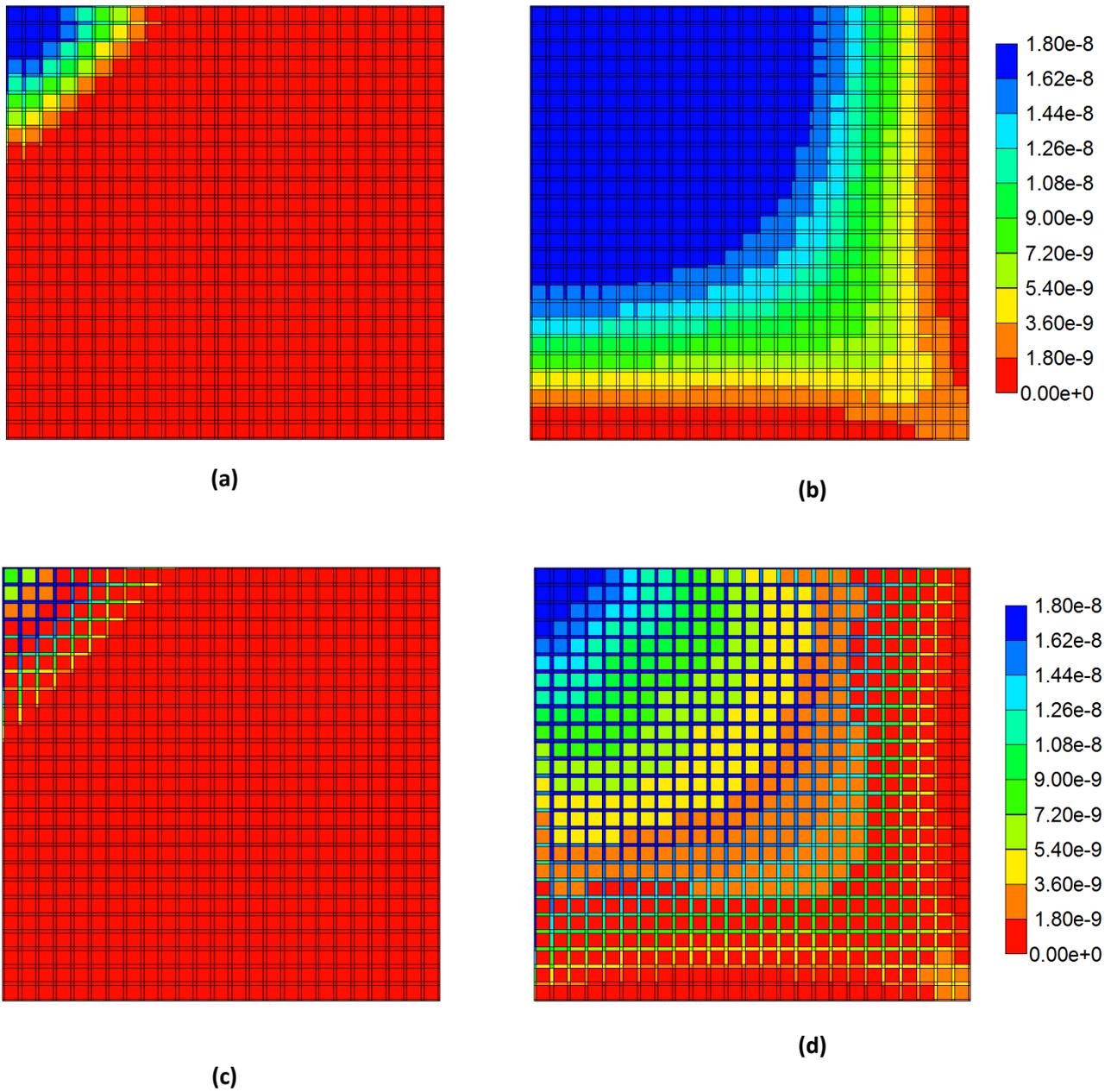

Figure 5: Non-reactive PAC profiles across the lobule (a) at 0.01 min, with diffusion effect, (b) at 0.14 min, with diffusion effect, (c) at 0.01 min, no diffusion effect, (d) at 0.14 min, no diffusion effect.

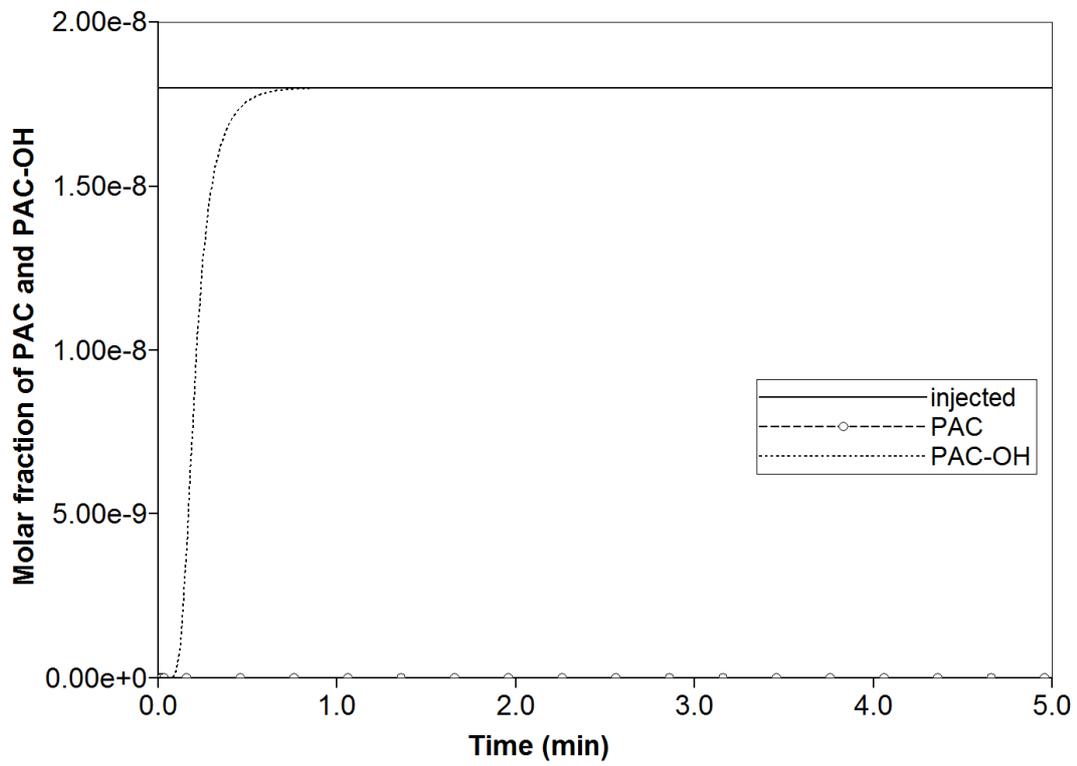

Figure 6: Reactive PAC and PAC-OH drug propagation across the lobule, with diffusion effects and base case metabolism.

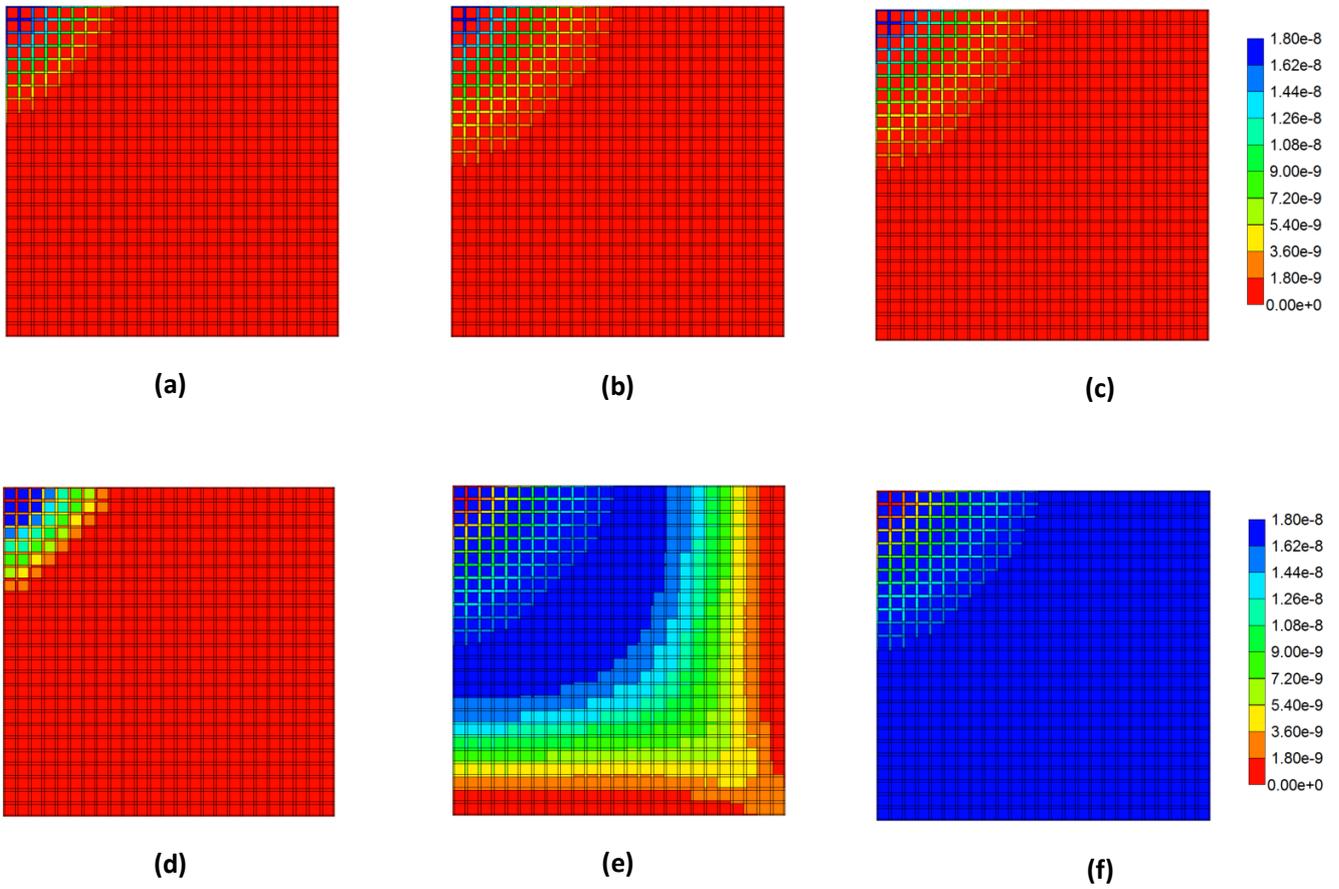

Figure 7: Reactive PAC and PAC-OH profiles across the lobule with diffusion effects and base case metabolism (a) PAC at 0.01 min, (b) PAC at 0.14 min, (c) PAC at 0.50 min, (d) PAC-OH at 0.01 min, (e) PAC-OH at 0.14 min, (f) PAC-OH at 0.50 min.

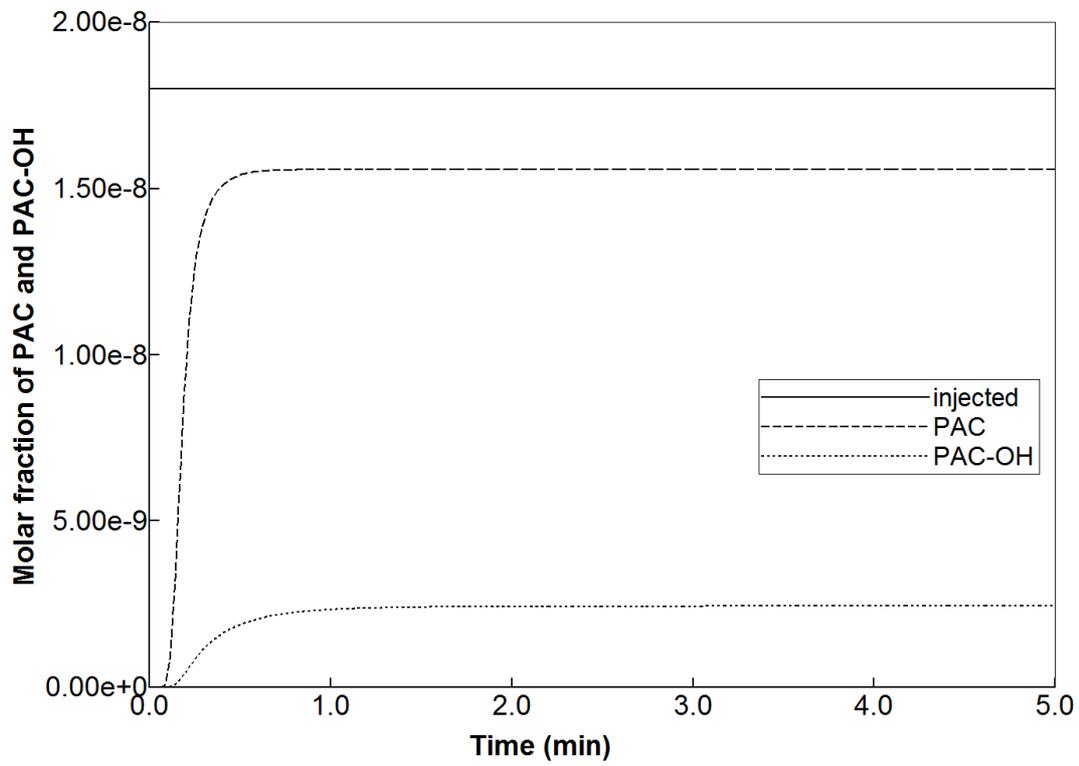

Figure 8: Reactive PAC and PAC-OH drug propagation across the lobule, without diffusion effects and base case metabolism.

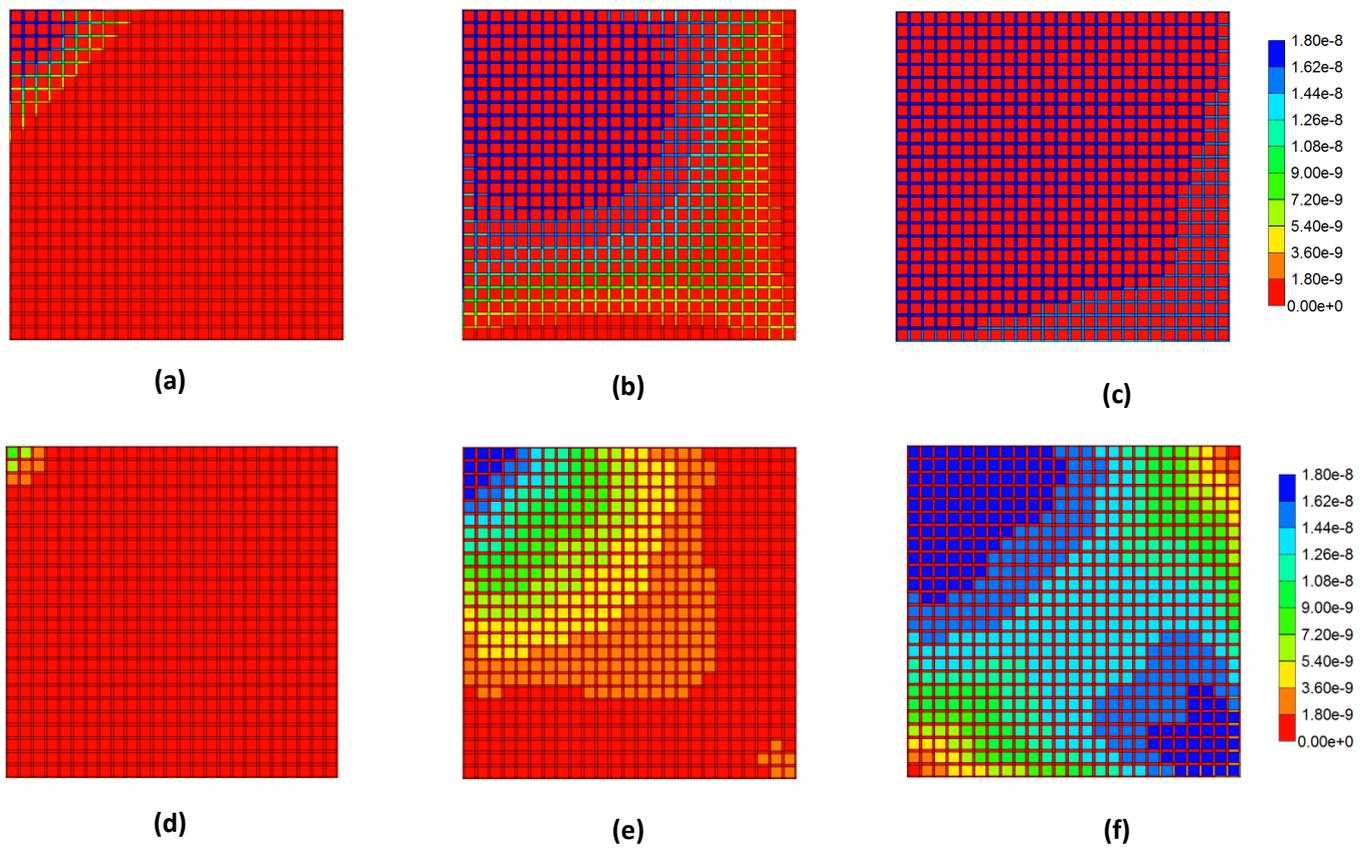

Figure 9: Reactive PAC and PAC-OH profiles across the lobule without diffusion effects and base case metabolism (a) PAC at 0.01 min, (b) PAC at 0.14 min, (c) PAC at 0.50 min, (d) PAC-OH at 0.01 min, (e) PAC-OH at 0.14 min, (f) PAC-OH at 0.50 min.

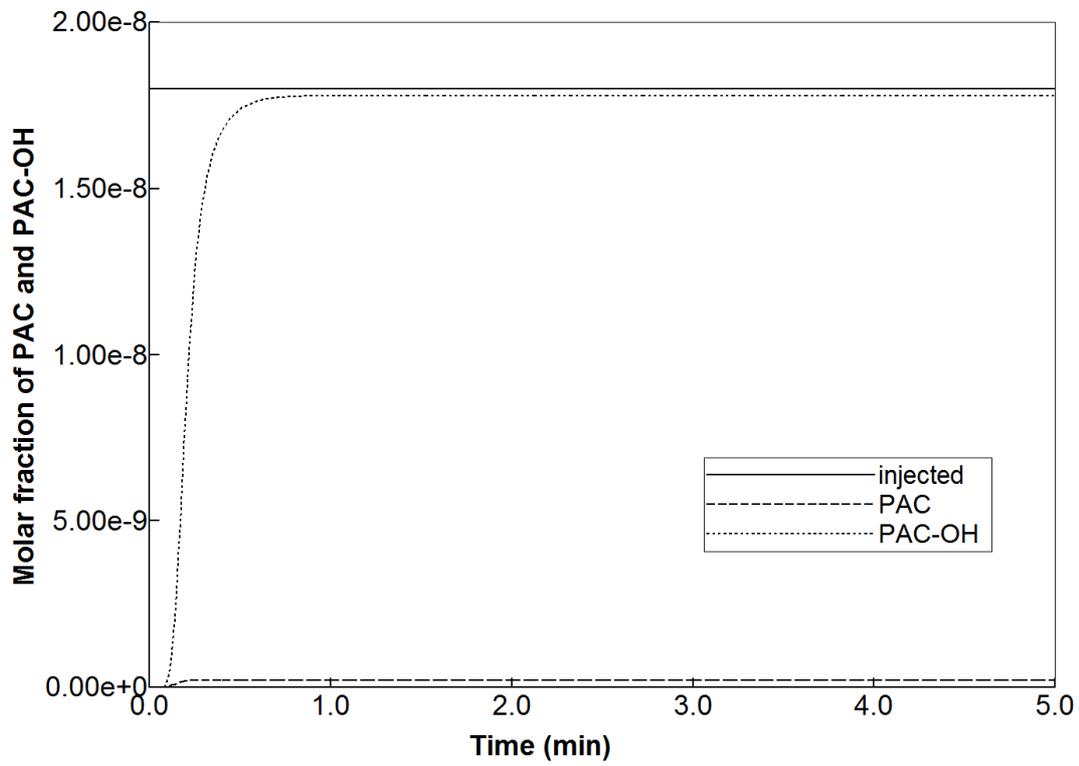

**Figure 10:** Reactive PAC and PAC-OH drug propagation across the lobule, with diffusion effects and 100-fold reduced metabolism.

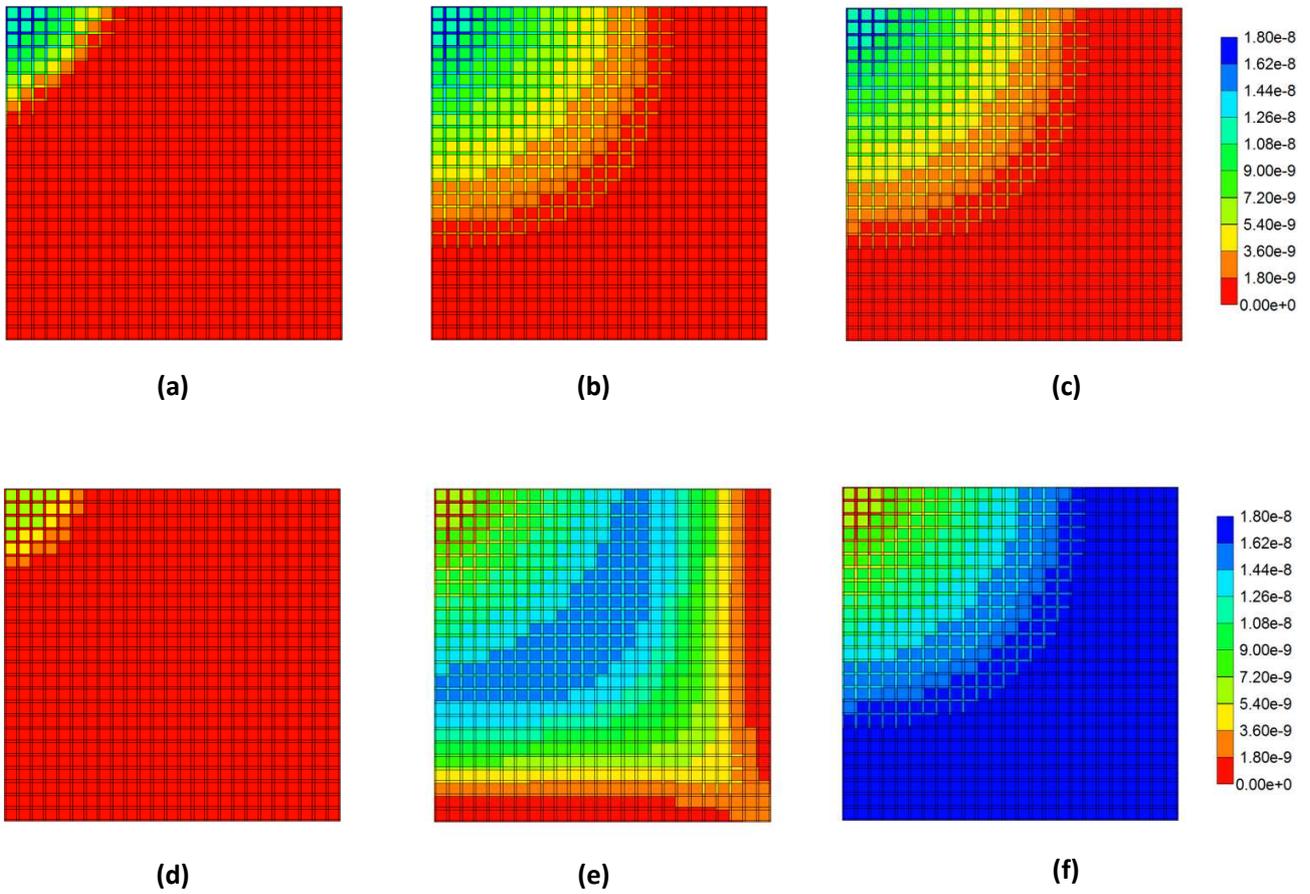

Figure 11: Reactive PAC and PAC-OH profiles across the lobule with diffusion effects and 100-fold reduced metabolism (a) PAC at 0.01 min, (b) PAC at 0.14 min, (c) PAC at 0.50 min, (d) PAC-OH at 0.01 min, (e) PAC-OH at 0.14 min, (f) PAC-OH at 0.50 min.

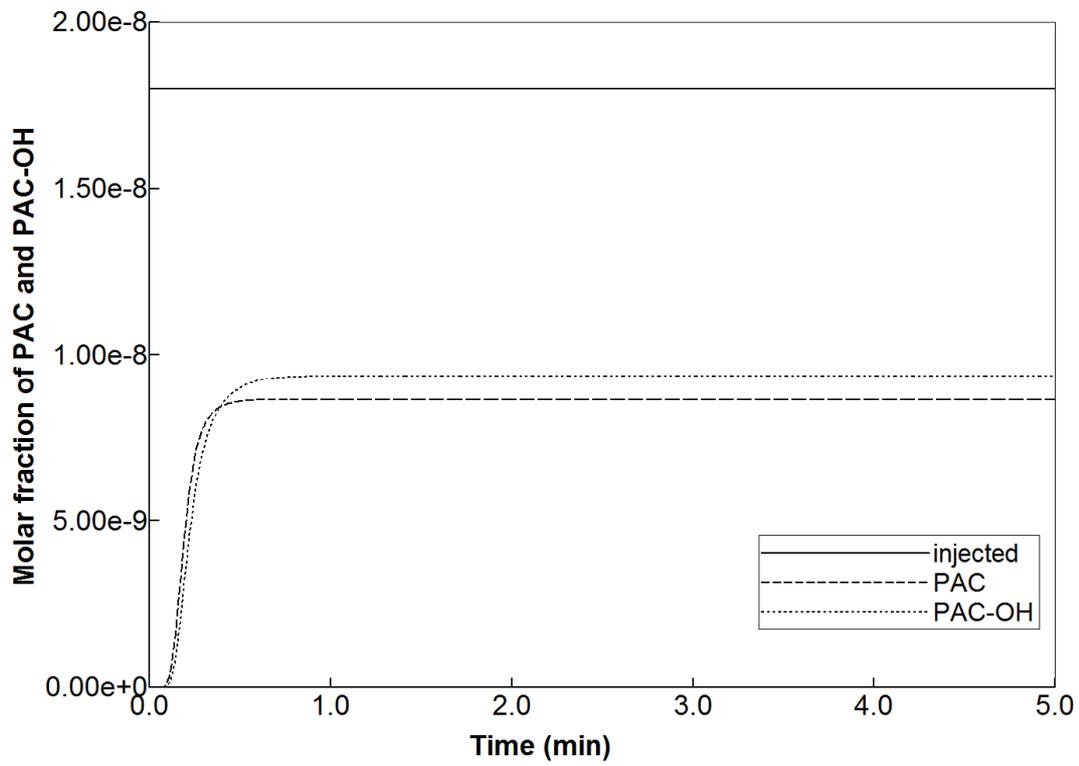

**Figure 12:** Reactive PAC and PAC-OH drug propagation across the lobule, with diffusion effects and 1000-fold reduced metabolism.

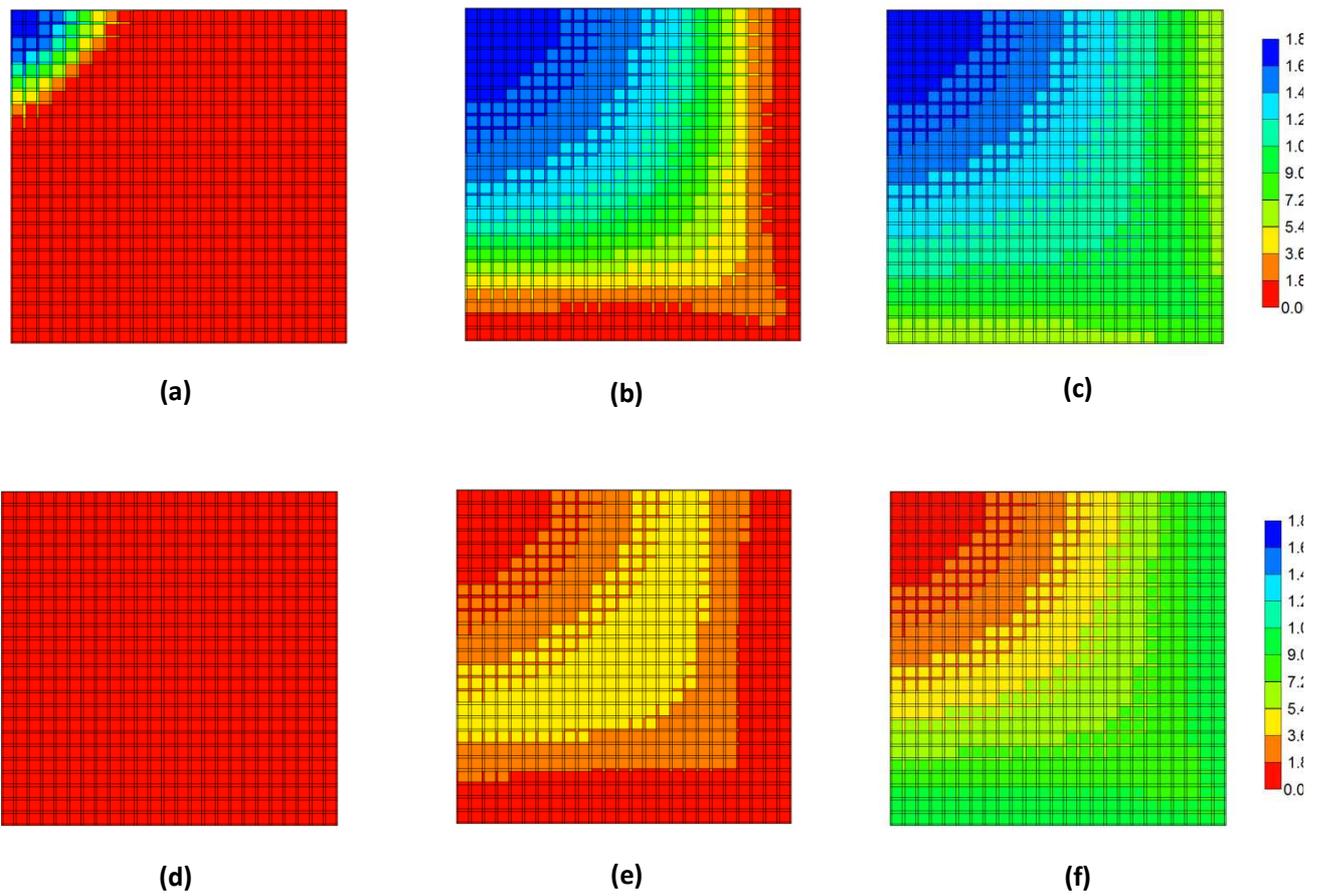

**Figure 13: Reactive PAC and PAC-OH profiles across the lobule with diffusion effects and 1000-fold reduced metabolism (a) PAC at 0.01 min, (b) PAC at 0.14 min, (c) PAC at 0.50 min, (d) PAC-OH at 0.01 min, (e) PAC-OH at 0.14 min, (f) PAC-OH at 0.50 min.**

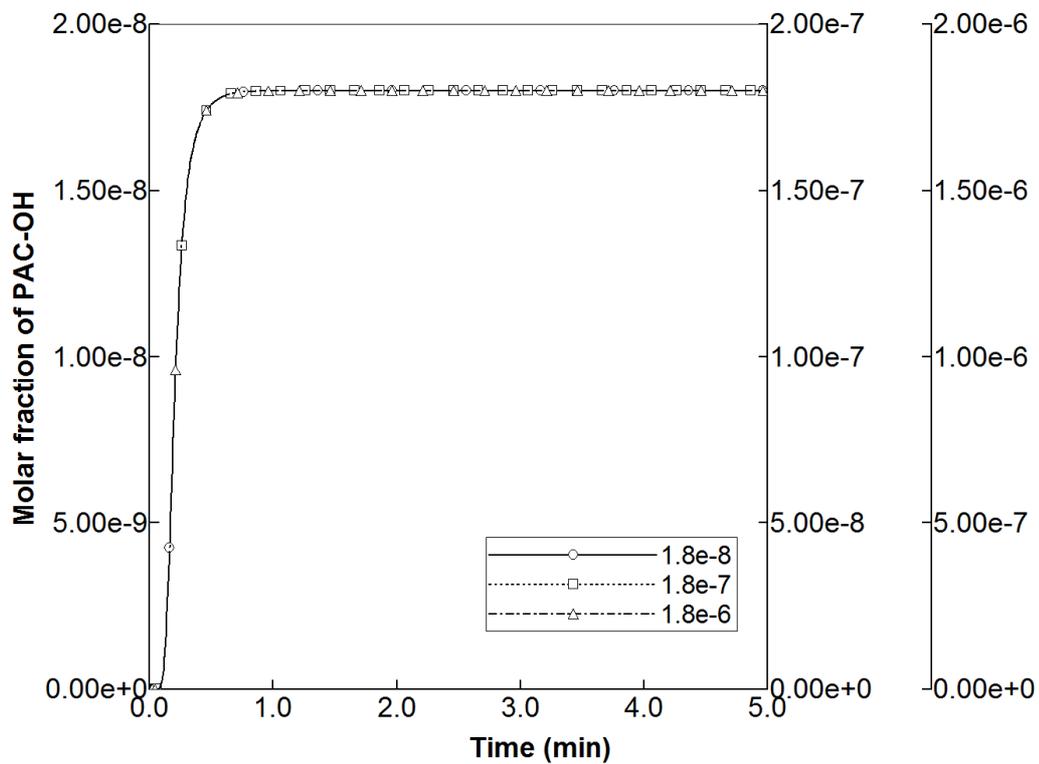

**Figure 14: PAC-OH metabolite production levels from various injected PAC concentrations.**